\documentclass[10pt,journal,onecolumn]{IEEEtran}
\IEEEoverridecommandlockouts
\usepackage{cite}
\usepackage{amsmath,amssymb,amsfonts}
\usepackage{algorithmic}
\usepackage{graphicx}
\usepackage{textcomp}
\usepackage{xcolor}
\usepackage{csquotes}
\usepackage[bookmarks=false]{hyperref}
\usepackage{tabularx,multirow,booktabs,blindtext}
\def\BibTeX{{\rm B\kern-.05em{\sc i\kern-.025em b}\kern-.08em
    T\kern-.1667em\lower.7ex\hbox{E}\kern-.125emX}}
\begin{document}

\title{Cloud as an Attack Platform\\
}

\author{\IEEEauthorblockN{Moitrayee Chatterjee\IEEEauthorrefmark{1}, Prerit Datta\IEEEauthorrefmark{1}, Faranak Abri\IEEEauthorrefmark{1}, Akbar Siami Namin\IEEEauthorrefmark{1}, and Keith S. Jones\IEEEauthorrefmark{2}}
\IEEEauthorblockA{\IEEEauthorrefmark{1} \\
\textit{Department of Computer Science}, \IEEEauthorrefmark{2} 
\textit{Department of Psychological Sciences}\\
\textit{Texas Tech University}\\
\{moitrayee.chatterjee, prerit.datta, faranak.abri, akbar.namin, keith.s.jones\}@ttu.edu}
}

\maketitle

\begin{abstract}

We present an exploratory study of responses from $75$ security professionals and ethical hackers in order to understand how they abuse cloud platforms for attack purposes. The participants were recruited at the Black Hat and DEF CON conferences. We presented the participants' with various attack scenarios and asked them to explain the steps they would have carried out for launching the attack in each scenario. Participants' responses were studied to understand attackers' mental models, which would improve our understanding of necessary security controls and recommendations regarding precautionary actions to circumvent the exploitation of  clouds for malicious activities. We observed that in $93.78\%$ of the responses, participants are abusing cloud services to establish their attack environment and launch attacks. 
\end{abstract}

\section{Introduction}

Cloud computing is an emerging computing paradigm that enables businesses and individuals to access computing resources as a service. These attractive features of cloud computing have gained significant attention from cyber attackers, and it is now extensively been exploited by adversaries for launching ``\textit{stealth}'' attacks or even setting up phishing Websites \cite{c15}. The study reported in this paper, alongside the security reports from various public cloud providers \cite{c2}, highlights that the cloud is continuously being \textit{weaponized} for launching attacks. According to the 2017 Microsoft Security Intelligence Report \cite{c2}, attackers can \enquote{weaponize} the cloud to create their own Virtual Machines (VMs) or gain access to or compromise other VMs. It has been reported that the Google Cloud Platform (GCP) has been abused for launching Denial of Service attacks or intrusion attacks\footnote{https://www.gcppodcast.com/post/episode-47-cloud-abuse-with-swati-and-emeka}. According to a report \cite{c3}, the \enquote{abuse and nefarious} use of {\it Infrastructure-as-a-Service} (IaaS) is one of the most critical security concerns in the cloud. Hackers are attracted IaaS because they can create computing accounts on cloud services with false identities, which enables them operate in obscurity.

This paper reports the findings from interviews conducted with 75 security professionals and ethical hackers who participated in the DEF CON and Black Hat professional hacking conferences. 
We observed that these professional hackers often employ common strategies to abuse the cloud platform for its resource-efficient features in order to remain stealth and silent while probing target machines, collecting victim data, discovering vulnerabilities, and launching attacks. This paper makes the following key contributions:


\begin{itemize}
    \item A generalization of cumulative interview data about how cyber attackers utilize cloud platforms for setting up their attack environment while remaining stealth. 
    \item Several recommendations for how cloud service providers can impede the misuse of their platforms, drawn from enumerating cyber attackers' generic attack steps. 
\end{itemize}
\section{Dynamic Model of Abuse Patterns}
\label{sec:recommend}
\begin{figure}
\begin{center}
  \includegraphics[width=0.7\linewidth, height=6.5cm]{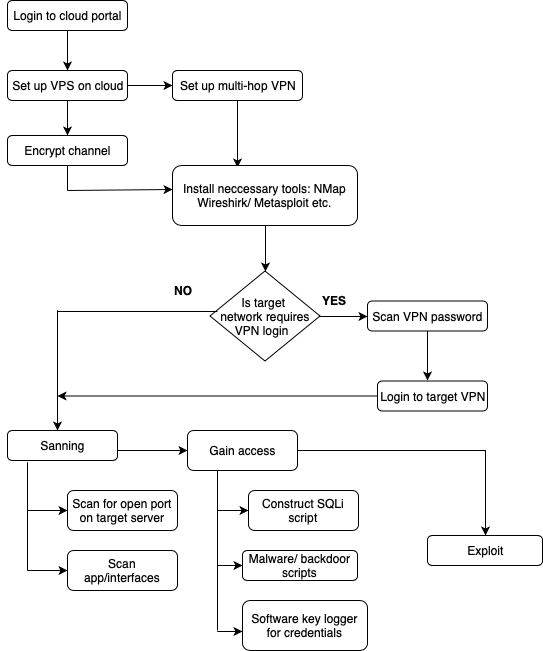}
  \caption{A dynamic model and activity-flow diagram for cloud abuse.}
  \label{fig:platform}
  \vspace*{-0.35in}
\end{center}
\end{figure}

As part of a larger project, in which one of the research objectives was to understand and analyze the attackers’ mental models when launching cyber attacks, our research team conducted a series of face-to-face and open-ended interviews with 75 professional hackers and penetration testing experts. We collected the interview responses interactively, and transcribed them into use cases during analysis.
One of the major observations from those interview responses was the abuse of cloud platform to set up the attack environment. We deduced the common patterns of cloud abuse and presented them as a dynamic model of the cloud-based attacks through an activity-flow diagram in Figure \ref{fig:platform}. \\ 
The Figure \ref{fig:platform} illustrates the common steps an attacker follows to abuse the cloud and performs the exploration, probing, and enumeration for constructing and transmitting malicious payloads to the target in order to launch attacks. The figure depicts the common ground patterns of use cases for creating a VM on IaaS cloud model, setting it up for performing reconnaissance, scanning, and gaining access to the target and launching an exploit, while remaining untraceable throughout various phases of the attack. The participants explained that they could set up a VPS (Virtual Private Server), a multi-hop VPN (Virtual Private Network), or encrypt the communication channel on the cloud VM (virtual machines). They could then install the necessary tool sets on the VM. By running all these tools using the computing resources of the VM on the cloud, attackers could craft and launch a SQL-injection attack, or propagate malware or run malicious scripts on a target machine, or even install software like a \textit{keylogger} on the target machine to obtain credentials. 

\section{Recommendations}
\label{sec:recommend}

The Google Cloud Platform has adequate tracking methods to monitor if any of the VMs are running any suspicious processes to circumvent the resources or network quota\footnote{https://cloud.google.com/compute/quotas}. AWS has GuardDuty \cite{c12} to trace malicious activities on the cloud, alongside Amazon Security Inspector \cite{c8} to perform security assessment. However, from our participants' data, we learned that attackers are erudite enough to evade the prevailing security measures on the cloud. In this section, we enumerate possible countermeasures and mitigation strategies to minimize the likelihood of cyber attackers abusing the IaaS cloud. The recommendations concern three aspects of an automated and effective defense mechanism including: 1) Prevention, 2) Detection, and 3) Recovery. 
\vspace{-0.1in}
\subsection{Prevention}
The prevention recommendations are to impede the abuse of the IaaS model in the first place. The following technical and regulatory actions could help prevent attackers from setting up their environments to abuse the cloud.

\noindent \textbf{Account Authentication.} To create a cloud account, an attacker needs only valid credit card information. However, several Websites offer fake credit card numbers\footnote{https://www.getcreditcardinfo.com}, so it is easy for attackers to create cloud accounts anonymously.  Thus, a thorough background check should be employed before activating a cloud account. Multi-factor authorization might also make attackers efforts more difficult. 

\noindent \textbf{Tracking Multi-hop VPN.} Cyber attackers frequently reported use  of multi-hop VPNs, which require significant network bandwidth. Thus, the amount of network bandwidth can serve as an indicator to detect suspicious activities. So, tracking the network quota and performing proactive monitoring when certain VMs start to exhaust the network quota can help to detect suspicious accounts. 

\noindent \textbf{Setting Up Firewalls and Update Software.} Cloud providers can enforce use of firewalls and encourage updating all software on VMs to latest security patches in order to protect against known vulnerabilities. 

\noindent \textbf{Trusted Software Repositories.} Cyber attackers need certain tools for scanning and reconnaissance activities. Cloud providers can enforce downloading of software tools from a trusted repository in order to prevent the use of adversarial tools on their platforms.

  \vspace{-0.1in}
\subsection{Detection}
Detection-based approaches should be implemented as a complimentary step to prevention-based approaches. That way, abusive activities can be identified as they take place.\\ 
\textbf{Proactive Forensic Analysis.} Public cloud providers can employ automated, periodic, and randomized forensic analysis of the Virtual Hard Drives (VHDs) to identify suspicious accounts. VHDs are the virtualized equivalent of the hard drives of VM instances. VHDs contain information on OS, files and folders, and processes. 

\noindent \textbf{Anomaly Detection.} Cloud platforms can benefit from utilizing automated anomaly detection tools and techniques in order to detect any suspicious activities in real time.  

   \vspace{-0.07in}
\subsection{Recovery}
Once an IaaS cloud instance is abused to launch an attack, it is important to ascertain  (1) the ways it was abused and (2) reinstate the system states (both attacker and victim):

\noindent \textbf{Blocking Malicious Traffic.} Setting up network rules to block any outgoing traffic from the attack VM.

\noindent \textbf{Isolating VMs.} Perform forensic analysis of the VHDs. 

\noindent \textbf{Enforce Blacklisting.} Identify the users of the accounts that perform malicious activity and blacklist them.

\section{Conclusion and Future work}


\label{sec:conclude}
This study highlights the needs for further research on how to prevent, detect, and recover from cyber attacks through cloud platforms. The detection methodologies presented here heavily depend on monitoring VM instances. Cloud service providers such as Amazon implement the Shared Responsibility Model\footnote{https://aws.amazon.com/compliance/shared-responsibility-model}, which makes users of cloud resources responsible for the safety measures \textit{inside} the cloud; whereas, the provider is responsible \textit{for} the cloud. The users of such platforms can programmatically enable alerts to identify various abuses\footnote{https://aws.amazon.com/blogs/mt/automating-processes-for-handling-and-remediating-aws-abuse-alerts/}, essentially making the users accountable for use or abuse of cloud.
The study points to the need for further work on developing a forensics suite and security testing framework \cite{c20} for cloud platforms. In an analogous way to zero-day malware \cite{c30}, it is also important to detect ``{\it zero-day abuse}'' of cloud. 
   \vspace{-0.03in}
\section*{Acknowledgment}
This research work is supported by National Science Foundation under Grants No: 1516636, 1723765, 1821560.

\vspace{12pt}

\end{document}